\documentclass[twocolumn,preprintnumbers,amsmath,amssymb,superscriptaddress]{revtex4}

\usepackage{graphicx}
\usepackage{dcolumn}
\usepackage{bm}
\usepackage{bbm}
\usepackage{amsmath}
\usepackage{amsthm}
\usepackage{dsfont}
\usepackage{amssymb}
\usepackage{enumerate}
\usepackage{epstopdf}

\newcommand{\bra}[1]{\langle #1|}
\newcommand{\ket}[1]{|#1\rangle}
\newcommand{\braket}[2]{\langle #1|#2\rangle}
\newcommand{\ketbra}[2]{|#1\rangle\!\langle#2|}
\newcommand{\id}{\mathbbm{1}}

\newtheorem{defin}{Definition}

\newcommand{\trans}{^\mathrm{T}}
\newcommand{\expt}[1]{\langle #1 \rangle}

\makeatletter
\newcommand*{\balancecolsandclearpage}{%
  \close@column@grid
  \clearpage
  \twocolumngrid
}
\makeatother

\let\oldmarginpar\marginpar
\renewcommand\marginpar[1]{\-\oldmarginpar[\raggedleft\marginparsize #1]%
{\raggedright\marginparsize #1}}

\usepackage{color}
\usepackage[dvipsnames]{xcolor}

\theoremstyle{plain}

\newtheorem*{theorem*}{Theorem}

\theoremstyle{definition}

\begin{document}

\setlength{\tabcolsep}{1ex}

\title{How uncertainty enables non-classical dynamics}

\author{Oscar~C.~O.~Dahlsten}
\affiliation{Atomic and Laser Physics, Clarendon Laboratory,
University of Oxford, Parks Road, Oxford OX13PU, United Kingdom}
\affiliation{Center for Quantum Technologies, National University of Singapore, Republic of Singapore}

\author{Andrew~J.~P.~Garner}
\affiliation{Atomic and Laser Physics, Clarendon Laboratory,
University of Oxford, Parks Road, Oxford OX13PU, United Kingdom}

\author{Vlatko~Vedral}
\affiliation{Atomic and Laser Physics, Clarendon Laboratory,
University of Oxford, Parks Road, Oxford OX13PU, United Kingdom}
\affiliation{Center for Quantum Technologies, National University of Singapore, Republic of Singapore}

\date{\today}

\begin{abstract}
The uncertainty principle limits quantum states such that when one observable takes predictable values there must be some other mutually unbiased observables which take uniformly random values.
We show that this restrictive condition plays a positive role as the enabler of non-classical dynamics in an interferometer. 
First we note that instantaneous action at a distance between different paths of an interferometer should not be possible. We show that for general probabilistic theories this heavily curtails the non-classical dynamics.
We prove that there is a trade-off with the uncertainty principle, that allows theories to evade this restriction.
On one extreme, non-classical theories with maximal certainty have their non-classical dynamics absolutely restricted to only the identity operation.
On the other extreme, quantum theory minimises certainty in return for maximal non-classical dynamics.
\end{abstract}

\maketitle

{\bf Introduction.}
The uncertainty principle stipulates that if the outcome of some observable of a quantum system is predictable, there will be another observable which must be unpredictable. It is a quintessentially quantum phenomenon, as in classical probability theory there is no ban on systems where all quantities can be deterministically known, and has been the subject of much discussion since the early days of quantum theory~\cite{Heisenberg30,Bohr58}.

A deeper understanding of this principle is a key aim in quantum foundations, believed to be holding the key to the understanding of a wide host of quantum phenomena. 
One important insight is that one may formulate theories similar to quantum theory, with the crucial difference that there are measurements that cannot be measured at the same time, but that are not subject to an uncertainty relation\cite{PopescuR94,Barrett07}. 
These theories can, as a direct consequence of having less (or even no) uncertainty, have more Bell violation than possible in quantum theory and may allow for greater work extraction than permitted by the second law of thermodynamics~\cite{PopescuR94,VerSteegW08,OppenheimW10,HanggiW13}. 
In these cases the uncertainty principle acts as a fundamental \mbox{{\em limiting} factor.}

Here, we show that actually the uncertainty relation also has a very positive {\em enabling} effect; in all probabilistic theories where immediate action at a distance is impossible, there is a trade-off between the amount of non-classical dynamics on the one hand and the amount of uncertainty on the other. The mathematical argument for this is very similar to how a body sitting on a surface with total friction is enabled to spin around if its shape is restricted such that it only has one point on the surface: sometimes a restriction on one feature can reduce restrictions on others. 
Quantum theory, we show, maximises the amount of non-classical dynamics, by maximising the uncertainty restriction. In the other extreme, if one has incompatible measurements with no uncertainty relation at all, then {\em any} state-changing transformation (whether reversible or irreversible) violates the restriction of no action at a distance. 
In this sense, uncertainty plays a positive role as the enabler of non-classical dynamics in quantum theory. 

We formalise the idea of probabilistic theories in general through the convex framework for probabilistic theories ~\cite{Hardy01, Mana03, BarnumBLW06, Barrett07, DakicB11, MasanesM11}, which has been used to show interesting results in the fields of information theory~\cite{Barrett07}, statistical mechanics~\cite{MullerDV11} and axioms for quantum theory~\cite{Hardy01, DakicB11, MasanesM11}. 
As an operational framework, any experiments yielding tables of data can be described in this way~\cite{Hardy01, Mana03}.

We use the term `non-classical dynamics' as a generalisation of what in quantum theory would be called `phase transformations'. These include the transformations that change the phases associated with different branches of an interferometer~\cite{GarnerDNMV13}. 

\begin{figure}[tb]
\centering
\includegraphics[width=85mm]{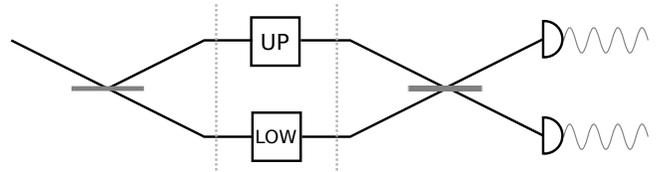}
\caption{ {\bf The Mach-Zehnder interferometer}: Branch locality, the restriction whose consequences the uncertainty principle enables escape from, states that if the (possibly post-quantum) particle is with probability 1 to be found in one of the branches of the interferometer, then operations on the other spatially disjoint branch cannot change the (operational) state of the particle. } 
\label{fig:interferometer}
\end{figure}

We here identify a particular type of `no action at a distance'-principle by considering a system which may be found in certain spatial regions with some probability.
We demand that if this system has no probability of being found in a particular region, then actions on that region cannot have an observable impact on the system. We call this innocent-sounding principle {\em branch} locality, because we are envisaging different branches of an interferometer (see Figure~\ref{fig:interferometer}). 
This statement might at first glance appear tautological, but we show that without uncertainty it has dire consequences for the allowed dynamics of the system.
(For discussions relating to interference in post-quantum theories in a different context see~\cite{Sorkin94, BarnettDR08, UdudecBE11}.)

We proceed by introducing the key concepts of the operational framework for probabilistic theories, such as states and transformations, as well as examples of theories, including quantum theory and the so-called box-world theory. We do this in the context of an interferometer, which is the physical scenario we focus on here. Then we define branch locality in that language. We show how in box-world, which has no uncertainty, no state-changing transformations at all can exist without violating branch locality. We then consider why the proof does not go through in quantum theory, showing that this is due to the uncertainty relation. We consider the theories in between these two extreme cases, theories with intermediate amounts of local dynamics, and we show that the amount of uncertainty regulates the amount of local dynamics.  Finally we discuss the implications, in particular with regards to computation with different non-classical theories.


{\bf Describing interferometers in the convex probabilistic framework.} 
We start with the simplest quantum case and then generalise it.
In the case of quantum theory one may describe an ideal Mach-Zehnder interferometer (Figure~\ref{fig:interferometer}) using a single qubit, i.e.\ a 2-dimensional Hilbert space. 
The state after the first beam-splitter can be expressed in the which-branch basis as  $\ket{\psi}=c_{\mathrm{up}}\ket{z_{\mathrm{up}}}+c_{\mathrm{low}}\ket{z_{\mathrm{low}}}$. 
The observable giving the expected position corresponds to $Z=z_{\mathrm{up}}\ket{z_{\mathrm{up}}} \bra{z_{\mathrm{up}}}+z_{\mathrm{low}}\ket{z_{\mathrm{low}}} \bra{z_{\mathrm{low}}}$ for some labels $z_{\mathrm{up}}$, 
$z_{\mathrm{low}}$ that we assign to the respective branches. Here, it will be convenient to label these $\pm 1$ respectively so that the observable is modelled by the Pauli-matrix $Z=\ket{z_{\mathrm{up}}} \bra{z_{\mathrm{up}}}-\ket{z_{\mathrm{low}}} \bra{z_{\mathrm{low}}}$. (The argument also works for other labellings than $\pm 1$.)

\begin{figure}[ht]
\centering
\includegraphics[width=70mm]{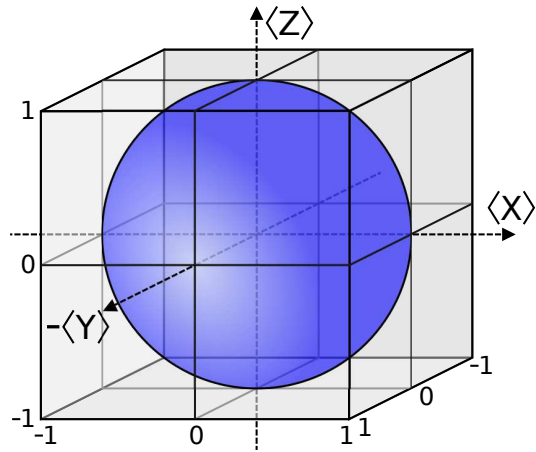}
\caption{{\bf State-spaces}: the quantum state-space is the (Bloch) sphere. 
We also consider the possibility of other states outside the sphere. 
The case of the maximal cubic state-space is an instance of the so-called `box-world'. 
The branch locality restriction mandates that states on the upper and lower plane are invariant under any transformation.
The Bloch sphere, respecting the uncertainty principle, touches the cube at only one point on each face and is unrestricted in its dynamics, but the box-world cube is totally frozen.
} 
\label{fig:phase_space_3D}
\end{figure}

The state-space of a qubit can be represented by real vectors using the well-known Bloch sphere, see Figure~\ref{fig:phase_space_3D}. 
Here a state is represented by a real-numbered vector of expectation values: $\left[ \langle X \rangle,\, \langle Y\rangle ,\, \langle Z\rangle \right]\trans$ where $X$ and $Y$ are the other two Pauli-matrices, and $\langle g \rangle=p(g\!=\!+1)-p(g\!=\!-1)$. Mixtures of states correspond to probabilistic (convex) combinations of these states, lying inside the sphere of pure states defined by 
\begin{equation}
\label{eq:sphere}
\langle X \rangle^2 + \langle Y \rangle^2 + \langle Z \rangle^2 = 1. 
\end{equation}
The above equation constitutes an uncertainty relation; for example if $\langle Z \rangle=1$ one must have $\langle X \rangle=\langle Y \rangle=0$. The more familiar formulation in terms of standard deviations, that $\Delta X \Delta Z\geq \frac{1}{2}|\langle \left[X,Z\right] \rangle|=|\langle Y\rangle|$, is implied by Eq.~\ref{eq:sphere} (recall that $(\Delta g)^2=\left<g^2\right>-\left<g\right>^2$).    

The real vector used above amounts to an {\em operational} description of the state. 
One may now entertain the possibility of post-quantum states by associating them with points outside the sphere of pure quantum states. 
In the present representation of states, the state $(1,\, 1,\, 1)\trans$ for example is not allowed in quantum theory as it violates the uncertainty principle of Eq.~\ref{eq:sphere}. 
We shall here a priori allow such states and later rule them out. 
In fact we shall only a priory assume that the theory fits into the convex framework (essentially any experiment yielding a data-table can be described in this manner~\cite{Hardy01, Mana03}). 
A key rule is that the state can be represented as a real vector $\vec{s}$.
A measurement is associated with a set of outcomes, $\{e_i\}$, each also represented by a real vector $\vec{e_i}$ (known as an {\em effect}) such that the probability of each outcome $e_i$ for that measurement on a state $\vec{s}$ is given by the inner product,  $p(e_i)=\vec{e_i}\cdot\vec{s}$. 
We use the concept of a {\em maximal measurement}, which distinguishes as many pure states as possible in a single shot (a generalisation of a quantum projective measurement).

It will be crucial to our argument to consider {\em transformations} of states. 
These must take all allowed states to allowed states.
They must also respect the linearity of probabilistic mixtures: for a transformation $T$ acting on a mixture of two states, $\nu_1$ and $\nu_2$, we have  $T\left(p_1\vec{\nu_1}+p_2\vec{\nu_2}\right)=p_1T\left(\vec{\nu_1}\right)+p_2T\left(\vec{\nu_2}\right)$.  
These transformations are real-numbered matrices acting on state vectors, up to the subtlety that one should now add an extra component $n$ to the state vector corresponding to the `normalisation' of the state ($n=1$ for normalised states). 
Phase transformations have been recently defined in this framework\cite{GarnerDNMV13}, generalising the idea of a phase plate in quantum theory, and it was shown that a theory is classical (meaning it can be described as classical probability theory) if and only if it has non-trivial phase transformations with respect to a type of measurements called maximal (this is a generalisation of the idea of the idea of projective measurements). 
We shall therefore here refer to phase transformations as non-classical transformations.

A {\em theory} is specified by the set of states (which implicitly assumes a set of measurements and outcomes has been defined) and the allowed transformations. 
As well as making general statements about all theories, we shall refer to three concrete examples. 
The {\em quantum} qubit case has states represented as $\vec{\nu}= \left[n, \langle Z\rangle,  \, \langle X\rangle, \, \langle Y\rangle \right]\trans$. 
The allowed transformations consist of both reversible SO(3) transformations as well as linear transformations shrinking the sphere.
Secondly, we shall call the case of a diagonal density matrix (s.t. $\langle X\rangle=\langle Y\rangle=0$) the {\em classical} case, modelled as $\vec{\nu}= \left[n, \langle Z\rangle , \, 0, \, 0 \right]\trans$. 
Here any matrix preserving or shrinking the line of states is allowed.
Finally, the maximal state-space of all probabilistic mixtures of the corners $(n,\pm n,\pm n,...)$ is known as the state-space of a single system in {\em box-world}.
The special case of  $(n,\pm n,\pm n)$ is termed a {\em gbit}~\cite{Barrett07}. 
This can (for n=1) be visualised as the square X-Z plane slice of the cube in Figure~\ref{fig:phase_space_3D}.
Gbits are currently of great interest in the context of understanding whether there can be Popescu-Rohrlich (PR) boxes.
These are hypothetical maximally Bell-violating systems (see~\cite{PopescuR94}).
The pure states of a gbit, the corners of the maximal state-space, are the conditional marginal states of a PR-box\cite{Barrett07} in the same way that pure qubit states are related to Bell states.
Thus if a PR-box can exist then so can a gbit. 
In box-world the allowed transformations on single systems are normally taken to be any matrix that preserves or shrinks the state-space (but our arguments will apply even if one is not so permissive with the transformations).

{\bf Branch Locality restriction}. 
Branch locality, as described in Fig~\ref{fig:interferometer}, can now be formalised as an operational principle.

\begin{defin}[Principle of {\em branch locality}] Physical actions on one region of space have no immediate effect on systems with no probability of being detected in that region.
In particular, let the branch $b$ be associated with measurement-outcome vector $\vec{e_b}$. Let the system be in a state $\vec{\nu}_b$ such that $p(Z=b)=\vec{e_b}\cdot\vec{\nu}_{b} =1$. We demand that $\vec{\nu}_b$ is left invariant by any transformation $T_{b'}$ on other branches $b'\neq b$: 
\begin{equation}
\label{eq:BL}
T_{\mathrm{b'}}\vec{\nu}_{b}=\vec{\nu}_{b}.
\end{equation} 
\end{defin}
We shall also impose a more obvious condition on operations at different branches. 
We take the state to be described by someone without access to outcomes of any measurements performed on the local branches.
Local transformations $T$ acting on the respective branches must then not alter the statistics associated with the $Z$ measurement. 
In the case of just two branches, using the Bloch-sphere representation: if the transformation takes $\langle Z\rangle$ to $\langle Z\rangle'$ then this is simply written as  
\begin{equation} 
\label{eq:Zconserved}
\langle Z\rangle=\langle Z\rangle'.
\end{equation}
In the language of \cite{GarnerDNMV13} this amounts to demanding that the transformation is a phase transformation associated with the position measurement $Z$. 

We now show that {\em no transformations in box-world respect Branch locality}. 
Here in the main body we give a more pedagogical argument for the simplest case of box-world, corresponding to two branches, and in the technical appendix we prove this statement for the most general case of box-world. 

Using the notation defined above, if we take a state to be in the upper branch with certainty it must have the form  $\vec{\nu}_\mathrm{up} = \left[n, \, n,  \, \langle X\rangle \right]\trans$ (recall that $n$ is the state normalisation with $n=1$ for a normalised state, and so $\langle Z\rangle=n$ for the upper branch). 

Consider an operation on the lower branch. From the above considerations, Equation~\ref{eq:BL} and Equation~\ref{eq:Zconserved} both hold. 
Recalling moreover that the transformation is a real matrix, it follows that: 
\[T_\mathrm{low} \vec{\nu}_\mathrm{up} \! = \! \left[\! \begin{array}{ccc} 
a & b & c \\
d & e & f \\
g & h & i
\end{array}\! \right]\!\left[ \! \begin{array}{c} 
 n \\
 n \\
\! \! \langle X\rangle \!\! \\
\end{array} \! \right]\!=\!\left[\! \begin{array}{c} 
\! (a+b)n + c\langle X\rangle \!\! \\
\! (d+e)n + f\langle X\rangle \!\! \\
\! (g+h)n + i\langle X\rangle \!\! \\
\end{array}\! \right]\!\!=\! \! \left[\! \begin{array}{c} \label{eq:gbit}
n \\ 
n \\
\!\! \langle X\rangle \!\! \\
\end{array} \!\right]\!.\] 

Consider the ranges of the different variables for gbits: $n$ can take values in the range $0$ to $1$ and $\langle X\rangle$ as well as $\langle Z\rangle$ from -$n$ to $n$. 
Note that even when $\langle Z\rangle=\pm 1$, $\langle X\rangle$ is free to take any value in the range $-n$ to $n$.
Note also that $T_\mathrm{low}$ is independent of the state it acts on. 
It follows with a little work that  
\[ T_\mathrm{low} =\left[ \begin{array}{ccc} 
a & b  & c \\
d & e  & f \\
g & h  & i
\end{array} \right] =\left[ \begin{array}{ccc} 
1 & 0 & 0\\
0 & 1 & 0 \\
0 & 0 & 1 \\
\end{array} \right]. \]

Similarly by considering $\langle Z \rangle = -1$ we have $T_\mathrm{up}=\id$.
Thus we have shown that all dynamics violate branch locality in box-world.
We prove this statement for any number of branches in the {\em technical appendix}. 

{\bf Uncertainty principle vs. Branch Locality.} 
The above proof for a gbit does not carry through in quantum theory.
The proof makes use of that fact that, for gbits, when $\langle Z\rangle=\pm1$, $\langle X\rangle$ is still free to take all possible values. 
This is a violation of the quantum uncertainty principle and a key difference between box-world and quantum theory. 
One may therefore think that the uncertainty relation renders the restriction of branch locality trivial.
In this paper, we prove a statement to this effect. 
We begin with some definitions which we will use to state our main result; these may also be of independent interest.

\begin{defin}[Conditionally restricted theory]
\label{def:cond_rest}
A theory is said to be conditionally restricted if in the case when one maximal measurement gives one of its outcomes with probability 1, 
the number of degrees of freedom in the state is reduced by more than the number of outcomes of the measurement. 
A theory is said to be {\em fully} conditionally restricted if in this case the state is fully determined. 
\end{defin}

In quantum theory there is a notion of mutually unbiased measurements, with the different Pauli matrices, $X$, $Y$ and $Z$ being prototypical examples. 
For a set of mutually unbiased measurements, knowing the outcome of one measurement should not give any information about the outcome of another.
We generalise this notion to convex probabilistic theories.

\begin{defin}[Mutually Unbiased Measurements]
\label{def:mum}
A set of measurements are mutually unbiased if for any valid state $\vec{s}$ there exists at least one valid state $\vec{s'}$ such that any one of the measurements in the set has its outcomes permuted relative to $\vec{s}$ and the statistics of all the other measurements in the set are the same as in $\vec{s}$.
\end{defin}
This definition reduces to the standard definition of mutually unbiased bases (MUBs) in the quantum case (see {\em technical appendix}).
We can now define what we will mean by a quantum-like uncertainty relation. 

\begin{defin}[Quantum-like Uncertainty Relation]
When one maximal measurement is known, any mutually unbiased measurements are uniformly random. 
\end{defin}

With these definitions, we can now state our main theorem. 
\begin{theorem*}[Main]
Transformations in fully conditionally restricted theories, such as theories with a Quantum-like uncertainty relation, are not restricted by Branch Locality.  
Transformations in theories that are not conditionally restricted are completely restricted by Branch Locality ($T=\id$). 
\end{theorem*}

This statement is proven in the technical appendix. 
An intuitive understanding may be reached on why the {\em restriction} placed by uncertainty enables non-classical dynamics: branch locality places a joint restriction on the states and transformations, and the restriction on transformations is weakened by strengthening the restriction on the set of states. 

{\bf Discussion. }
In this paper, we have demonstrated the importance of the uncertainty principle for theories that are subject to locality requirements. 
The restriction placed by locality is so strong that without the uncertainty principle to mitigate its effects, absolutely no non-classical dynamics are admitted.
One may say that uncertainty is the sacrifice that quantum theory must make in order to maximise its non-classical dynamics.

Our introduction and consideration of the branch locality principle has more dramatic implications than previous results concerning the restricted dynamics in box-world~\cite{Barrett07,  ShortB10, GrossMCD10}.
We have ruled out any non-trivial dynamics, whether reversible or not, and by non-trivial we mean any transformation that is not the identity (whereas the word `trivial' in the title of~\cite{GrossMCD10} refers to the lack of correlating interactions). 
Moreover we show the same restriction holds for {\em any} convex theory that is not conditionally restricted, not just box-world.

As any computation has to be performed as an evolution of a physical system, our results can be interpreted as saying that computation using a (two or multi-branch) Mach-Zehnder interferometer is trivial unless the uncertainty relation holds. 
This experimental setting stands out as being the original setting in which quantum computation was conceived, with the Deutsch-Jozsa algorithm~\cite{DeutschJ92} arising naturally by considering what one can do with a quantum system in an interferometer. In~\cite{CleveEMM98} it was argued that, more generally, key quantum algorithms can be viewed as a three stage interferometer experiment: (i) prepare a superposition of different branches, (ii) apply different phase-shifts to different branches, and (iii) bring the branches together and make a measurement, yielding information about the phase shifts that were done. The apparent ability to prepare and individually address several inputs in stages (i) and (ii) is called quantum parallelism~\cite{DeutschJ92} (distinct from classical parallel computation). Our result suggests that an uncertainty relation is required to achieve this parallelism, directing the search for post-quantum theories with stronger computational power to those that respect the uncertainty relation, such as `systems with limited information content'~\cite{Zeilinger99, PaterekDB10, BruknerZ09}.

{\bf Acknowledgements.---}
We gratefully acknowledge discussions and correspondence (in chronological order) with Yoshifumi Nakata, Mio Murao, Markus M\"uller, \v{C}aslav Brukner, Mehdi Ahmadi, Anton Zeilinger, Matt Pusey and Jerry Finkelstein, as well as funding from the National Research Foundation (Singapore), the Ministry of Education (Singapore), the EPSRC (UK), the Templeton Foundation and the Leverhulme Trust. OD was regularly visiting Imperial College whilst undertaking this research.


\clearpage
\section*{TECHNICAL APPENDIX}
{\bf The general probabilistic theory (GPT) framework.}  
We first introduce the key concepts of the operationalist approach we will use: {\em the framework for general probabilistic theories (GPT)} also known as the convex framework. 
For a more detailed description of the framework see e.g.~\cite{Hardy01,MasanesM11}. 
The framework is operational in the sense that essentially any experiment producing a data table can be described in this way~\cite{Hardy01, Mana03}. 
For readers familiar with quantum theory it can also be helpful to think of the GPT framework as a generalisation of quantum theory.
In quantum theory there is a system which is prepared in a state $\rho$, determined by the preparation in question. 
There is a set of measurements one may do, each represented by a set of projection operators $\{\Pi_i\}_{i=1}^{\dim{\mathcal{H}}}$ (or more generally POVM elements). 
The operationally significant quantities, the probabilities of given outcomes are given by $p_i=Tr(\rho\Pi_i)$. Viewed more abstractly, the state is a vector $\vec{\rho}$ in the vector space of Hermitian operators. 
A projection operator is also such a vector, $\vec{\Pi}$ say. 
In other words we may pick a basis of the vector space and write $\vec{\rho}=\sum_i \xi_i \vec{e_i}$ and $\vec{\Pi}=\sum_j \nu_j \vec{e_j}$. 
(Examples of such bases are the Pauli operators and the pure state basis given by Hardy in~\cite{Hardy01}). 
Note that the coefficients of these expansions are {\em real}, so this is termed a real vector space. 
$Tr(\rho\Pi)$ is then the Hilbert-Schmidt inner product (for Hermitian matrices) which we may write as $\langle\vec{\rho},\vec{\Pi} \rangle$.

If the basis elements are chosen so that they are orthogonal with respect to the norm, and all having the same inner product $c$ with themselves, we see that $\langle\vec{\rho},\vec{\Pi} \rangle=\vec{\xi}\cdot \vec{\nu}c$ where the right-hand-side is the standard Euclidean norm. 
Thus we may represent a quantum state, measurements on it, and the resulting probabilities in terms of real vectors and the Euclidean norm.   

In the framework we more generally represent the state of a system $\vec{s}$ as a real vector, and the measurement-outcome pairs, called `effects' for historical reasons, as real vectors $\vec{e}$ (for example, in the quantum case such a vector could be associated with X=+1, where X is the Pauli X). 
The probability of the outcome associated with a given $\vec{e}$ is given by $\vec{s}\cdot \vec{e}$. 
Part of the specification of a given theory is specifying which states and effects are allowed. 
All convex combinations (mixtures) of allowed states are always allowed, written as $\vec{s}=\sum_i p_i\vec{s_i}$ (hence the name `convex framework'). 
A state is said to be {\em pure} if it is not a non-trivial mixture of other states, otherwise it is called {\em mixed}. 

Transformations are represented as real matrices acting on the state vector (following from the requirement of respecting mixtures, see e.g.~\cite{Hardy01}). 
They must take all allowed states to allowed states, but there may be further restrictions specified for a given theory in the framework. A transformation $T$ is termed {\em reversible} if its inverse $T^{-1}$ is also allowed in the theory. 

As well as quantum theory and theories contained therein (such as classical probability theory), one may also formulate a theory called {\em box-world} in this way.
Box world contains all states that do not violate non-signalling (that the reduced state of one system is invariant under operations on another)~\cite{Barrett07}.
The standard version of box-world assumes there are only two binary outcome measurements under considerations. 
We label these $X$ and $Z$ and the outcomes $\pm 1$ in analogy with quantum theory. 
A normalised state can then, as discussed below, be represented as $\vec{s}=[\langle X\rangle\,\, \langle Z\rangle]\trans$ and is any mixture of the four extremal states $[\pm 1 \,\, \pm 1]\trans$. 
The most general single system box-worlds are $m$-in $n$-out box-worlds, which mean that one selects one measurement setting from $m$ possible settings and obtain $n$-valued outcomes. 
In particular, the $3$-in $2$-out box-world is the most analogous to a qubit in quantum theory.

\textbf{\textit{Expectation value representation.}} 
For binary measurements, we find it makes the notation very simple and easy to visualise by focusing on the expectation values of measurements, along the lines of the example for box-world presented in Results. 
In what follows, we define a representation of states in terms of expectation values and relate it to the more standard representation in terms of probabilities, including showing that the transformations are matrices also in the new representation.

Consider the case of states described in terms of two fiducial measurements with binary outcomes (measurements are called fiducial if their statistics are sufficient to determine the state).
Such outcomes need not be normalised, but we require the sum of probabilities of both measurements to be equal.

Consider the {\em probability} representation of a state~\cite{Barrett07}:
\[\vec{s}:= \left[ \begin{array}{c}
 p(X=+1) \\
 p(X=-1)\\
 \hline
p(Z=+1) \\
 p(Z=-1)\\
\end{array} \right].\]
The following is the alternative (normalisation-including) {\em expectation} value representation:
\[\vec{\nu}:= \left[ \begin{array}{c}
 n \\
p(X=+1)-p(X=-1) \\
p(Z=+1)-p(Z=-1)\\
\end{array} \right]=
\left[ \begin{array}{c} 
 n \\
\langle X\rangle \\
\langle Z\rangle\\
\end{array} \right],
\]
where the normalisation $n=1$ if the state is normalised and $n<1$ if it is subnormalised. (For a subnormalised state we still use the notation $\langle g\rangle:=p(g=+1)-p(g=-1)$, implying the range $-n\leq \langle g\rangle \leq n$.)
 
If a transformation $T$ acts as a matrix on the state-vector in the probability representation, as it should if it respects mixtures, is it also a matrix in the expectation value picture? Suppose for the sake of argument that (i) There exists a fixed matrix $M$ such that $\vec{\nu}=M\vec{s}$ for all states,  (ii) The effective inverse matrix $M^{-1}$ also exists satisfying $M^{-1}M\vec{s}=\vec{s}\,\,\forall \vec{s}$ . 
Then we can write
$$MT\vec{s}=MTM^{-1}M\vec{s}=MTM^{-1}\vec{\nu}=\vec{\nu'},$$
where $\vec{\nu'}$ is the expectation representation state after the transformation. Thus we see that {\em if the two assumptions above hold then the state transformations by a matrix also in the expectation value picture.} Moreover these two assumptions do hold here, with (for example):
\[M\!=\!\left[ \begin{array}{cccc} 
1/2 & 1/2 & 1/2 & 1/2 \\
1 & -1 & 0 & 0 \\
0 & 0 & 1 & -1 \\
\end{array} \right],\,
M^{-1}\!=\!\left[ \begin{array}{cccc} 
1/2 & 1/2 & 0 \\
1/2 & -1/2 & 0 \\
1/2 & 0 & 1/2 \\
1/2 & 0 & -1/2 \\
\end{array} \right].\] 
The above argument naturally generalises to more measurements. 

Note also that we could have used a different label for the positions in the probability picture (i.e. not $\pm1$) and then mapped that into the expectation value picture using the same matrix as above. 
In this sense our argument does not depend on how we have labelled the two positions.  

\textbf{\textit{Minimal representation of states.}} 
Choosing a good representation of the states and transformations significantly aids the proof, and we shall therefore allow ourselves to introduce a third representation, intermediate between the expectation value and probability representations described above. We take a state in the probability representation and re-express it in the following way:
\[ \left[ \begin{array}{c}
p(Z=0)\\
\vdots \\
p(Z=\max _Z)\\
\hline 
p(X_1=0)\\
\vdots \\
p(X_1=\max _{X1})\\
\hline 
p(X_2=0)\\
\vdots \\
\end{array} \right]
\mapsto
\left[ \begin{array}{c}
 n \\
p(Z=0)\\
\vdots \\
p(Z=\max _Z-1)\\
\
p(X_1=0)\\
\vdots \\
p(X_1=\max _{X1}-1)\\
p(X_2=0)\\
\vdots \\
\end{array} \right],\]
where in the case of Z the different numbers are arbitrary labels for the different branches. 

For example in the case of two branches and two fiducial measurements (labelling $up=0$ and $low=1$), the state would be expressed as  
\[\vec{\nu}:= \left[ \begin{array}{c}
 n \\
p(Z=0)\\
p(X_1=0)\\
\end{array} \right].\]

Any state where all measurements have the same degree of normalisation can be expressed in this representation, and one sees that there exists a matrix that maps states from the probability representation to this new one, as well as another matrix for the other direction. Thus, by the arguments earlier in the appendix, matrices representing transformations in the probability picture are also matrices in this new picture. The advantage of this new picture over the probability picture for our purposes is that all parameters, for a given normalisation $n$, are independent. Moreover the advantage over the expectation value representation is that this representation is more easily generalised to any number of branches.

{\bf Mutually unbiased measurements.}
We can express the definition of mutually unbiased measurements (Definition~\ref{def:mum}) in more mathematical language, by first considering what it means for one measurement to be unbiased with respect to another.
Consider a measurement $X$ with effects $\{\vec{x_i}\}$ and measurement $Y$ with effects $\{\vec{y_i}\}$.
$X$ is unbiased with respect to $Y$ if for each permutation of the $X$ outcome statistics, there should exist a $P_x$ such that the total set of $X$ outcomes, $\{\vec{x_i}\cdot P_x \vec{s}\}$ is equal to $\{\vec{x_i}\cdot \vec{s}\}$, and none of the $Y$ outcomes are changed, such that $\vec{y_i}\cdot\vec{s}=\vec{y_i}\cdot P_x \vec{y_i}$ for all $\vec{y_i}$, and further more $P_x\vec{s}$ should be a valid state, and this should hold for all valid states $\vec{s}$ in the theory. (Note that we do not require $P_x$ to be a physically allowed transformation- just an automorphism of the state-space.)

We then say a set of measurements is {\em mutually unbiased} if each measurement in the set is unbiased with respect to every other measurement in that set. By this definition, a state-space expressed in representation of mutually unbiased measurements is symmetric under the relabelling of any of its measurement outcomes.

{\em Recovery of the quantum definition.---}
We consider whether this definition recovers the usual quantum definition of mutually unbiased bases: that for two measurements $X$ and $Y$ associated with $d$ eigenstates $\{\ket{x_i}\}$ and $\{\ket{y_i}\}$ respectively, $X$ and $Y$ are mutually unbiased if $|\braket{x_i}{y_j}|^2 = 1/d$ for all $i$ and $j$.

We consider some projective measurements $X$ and $Y$ with $d$ independent outcomes. It is possible to assume in quantum theory (e.g. following a measurement) that we are in an eigenstate with respect to one of these measurements, and therefore states of the form $\rho_i = \ketbra{x_i}{x_i}$ are valid states.

If we are in the eigenstate $x_j$ of $X$, the only allowed set of measurement statistics for the $Y$ measurement is given by $P(Y=y_i|X=x_j)=\braket{x_j}{y_i}\braket{y_i}{x_j} = |\braket{y_i}{x_j}|^2$.
If we permute our state from $\ket{x_j}$ to $\ket{x_k}$, again there should only be one allowed set of $Y$ statistics, given now by $P(Y=y_i|X=x_k) = |\braket{y_i}{x_k}|^2$.
By our definition, if $X$ and $Y$ are mutually unbiased, the state in which we have permuted $X$ without altering the $Y$ statistics must be allowed. 
As there is only one set of Y statistics for the state $\ket{x_k}$, we therefore see that $|\braket{y_i}{x_1}|^2 = |\braket{y_i}{x_2}|^2 = \ldots = |\braket{y_i}{x_d}|^2$ for each $i$.

Similar logic can be made for pure states of $Y$, such that $|\braket{x_j}{y_1}|^2 = |\braket{x_j}{y_2}|^2 = \ldots = |\braket{x_j}{y_d}|^2$ for each $j$.
As $|\braket{x_j}{y_i}|^2 = |\braket{y_i}{x_j}|^2$, this implies that this inner product squared is the same for every pair $\ket{x_j}$, $\ket{y_i}$.

Thus, as for a normalised Y measurement $\sum_j |\braket{x_i}{y_j}|^2 = 1$ for each pure state $\ket{x_i}$, then we can replace the sum with any element in the sum repeated $d$ times, such that $d|\braket{y_i}{x_j}|^2=1$ for all, $i$,$j$ and hence $|\braket{y_i}{x_j}|^2 = \frac{1}{d}$ for all $i$ and $j$, recovering the usual definition for $X$ and $Y$ to be mutually unbiased.
Furthermore, it should be noticed that in quantum physics, because each pure state is fully conditionally restricted in its choice of other measurement outcomes, any mutually unbiased basis will automatically respect an uncertainty principle.

{\em Theories without uncertainty.---}
We carefully note that a theory that can represented by mutually unbiased measurements does not necessarily have to obey an uncertainty relation.
A counter example of a theory represented by mutually unbiased measurements, but without uncertainty, is a gbit composed of binary-outcome measurements.
As there are no conditional restrictions, any statistically possible state is allowed, permuting one of the measurements whilst leaving the others unchanged is a reflection in an axis of the state-space, and this is an automorphism of the state-space, and so maps all valid states to valid states. 
This can be done for each measurement, and so the bases are mutually unbiased, even though uncertainty violating states such as $(\expt{X}, \expt{X}, \expt{Y}) = (1, 1, 1)$ are possible.

\vspace{1em}
{\bf Proof of main theorem.}
We restate our main claim for reference:

\begin{theorem*}[Main]
Transformations in fully conditionally restricted theories, such as theories with a Quantum-like uncertainty relation, are not restricted by Branch Locality.  
Transformations in theories that are not conditionally restricted are completely restricted by Branch Locality ($T=\id$). 
\end{theorem*}
\begin{proof}

We now consider theories that may incorporate many-branched interferometers, in which Z can take more than two possibilities, and there might be an arbitrary number of alternative $X$ measurements, each with an arbitrary number (greater than one) of possible outcomes.
In such theories where $Z$ measurement has $N$ possible outcomes, subject to the normalisation being fixed by $Z$ the total number of degrees of freedom from the other $X$ measurements is $M$, given by:
\begin{equation*}
M = \sum_i (\mathrm{outcomes}(X_i)-1). 
\end{equation*}
The total number of degrees of freedom in the state-space (and hence dimensions of the state vector in new representation we described above) is $d=N+M$. 
A transformation $\mathcal{T}$ on the state can therefore be represented by a $d\times d$ matrix.

If we fix $Z$ to take a definite outcome (that is $P(Z=z_j)=1$ for some $j$ and is $P(Z=z_i)=0$ for all $i\neq j$), then in general we have up to $M$ degrees of freedom available in our choice of $X$ statistics. 
The number of freedoms we have depends on whether the theory is {\em conditionally restricted} (recall definition~\ref{def:cond_rest}). In particular, we concentrate on the two extremal cases:
\begin{enumerate}[I.]
\item The {\em fully conditionally restricted} case, in which fixing $Z$ uniquely specifies our choice of $X$ statistics, such that there are no degrees of freedom left. 
Quantum theory, because of the constraint on its space space placed by the uncertainty principle, is in this category.
\item The {\em fully independent} case with no conditional restrictions, in which even after fixing $Z$, we still have the maximum number ($M$) of degrees of freedom in choosing a state. 
Box world is one theory which fits into this category.
\end{enumerate}

A fundamental result of linear algebra is that any matrix can be fully categorised by the complete set of its eigenpairs (eigenvectors with associated eigenvalues). 
We note that the branch locality restriction on a state $\vec{\eta}$ (when $\vec{\eta}$ has no probability of being in the branch where $\mathcal{T}$ is applied) has the form of an eigenvector equation on $T$ for solution $\vec{\eta}$ with associated eigenvalue $+1$:
\begin{equation*}
\mathcal{T}\vec{\eta} = \vec{\eta}.
\end{equation*}
Thus by applying such a restriction to generate a set of eigenpairs, it is natural that (at least partially) this restriction will characterise the nature of $\mathcal{T}$. 
We consider this in the two extremal cases:

{\em Fully conditionally restricted case---}
If the system is not in the branch $z_j$, there are $N-1$ other branches in which it could be instead.
For each state where the particle is definitely in one of these other branches, by the definition of {\em fully conditionally restricted} there is just one possible set of $X$ statistics. 
Therefore for each branch, we can only pick one allowed state $\vec{\eta}_i$, and this state must satisfy $\mathcal{T}\vec{\eta}_i=\vec{\eta}_i$.
There are $N-1$ of these independent states, corresponding to the $N-1$ possibly independent configurations of the $Z$ statistics.

{\em Fully independent case---}
Again, if the system is not in a particular branch there are $N-1$ different possible other branches the system could be in.
However for any of these other branches there are no longer any other constraints on the choice of $\eta$ in the equation $\mathcal{T}\vec{\eta} = \vec{\eta}$. 
For each branch, by the definition of a {\em fully independent} state, we can freely alter all $M$ degrees of freedom of $X$ in $\vec{\eta}$ and thus can fully span all possibilities of $X$ with a set of $M$ independent states satisfying the branch locality restriction. 
The set of independent state vectors from one branch will fully span all possible X statistics, and so each branch considered after this will only contribute one additional independent state vector. 
Taking into account that the branch locality restriction is applicable to $N-1$ branches, we see that these freedoms combine to give us $M+N-1$ independent state vectors associated with a $+1$ eigenvalue of $\mathcal{T}$.

Finally, having considered all other possibilities, we consider the states where the system is definitely in the branch on which the transformation is applied. Here, branch locality does not impose a restriction: the only restriction we have is that the statistics of $Z$ must be left undisturbed (i.e.\ that T is in the phase group of $Z$). 
In fixed theories, this trivially gives us another $+1$ eigenvector, as there is only one state $\vec{\eta}$ allowed when $Z$ is in a branch with certainty, and given that which branch this is can't change, the only state satisfying the restriction on $Z$ is $\vec{\eta}$ itself.
In fully independent theories, the argument is not so trivial; but rather we must rely on a property of stochastic matrices which states that any transformation has at least one state which is left invariant. 
Thus, for a transformation that only acts on the $X$ statistics, there will be at least one state $\vec{\eta}$ with the desired $Z$ statistics which is unchanged by the application of $\mathcal{T}$. 
In box-world, where any statistically possible state is allowed, it is clear that $\vec{\eta}$ will be a valid state. In general it could be that this state is forbidden for some reason; but this should not make it any less of a valid eigenvector for $\mathcal{T}$.

The eigenvector we find from either of these methods is evidently independent from any from the eigenvectors from branch locality requirement, as there is no convex combination of them which can yield the correct $Z$ statistics, and so we account for our final $+1$ eigenvector.

We note that if a $d \times d$ matrix has $d$ independent $+1$ eigenvalues, it {\em must be the identity matrix} (its diagonal form will be $\id$ and no change of basis will result in anything other than $\id$; expressing a general vector in terms of eigenvalues $\vec{v} = \sum c_i \vec{\eta}_i$, then $\mathcal{T}\vec{v} = \sum c_i \mathcal{T} \vec{\eta}_i = \sum c_i \vec{\eta}_i = \vec{v}$ and so every vector is left unchanged by $\mathcal{T}$, which is another definition of an identity operation).

We recap the total number of degrees of freedom accounted for in the two special cases:
\begin{enumerate}[I.]
\item {\em Fully conditionally restricted theories} must have at least $N$ eigenvectors with eigenvalue $+1$. 
This still allows us the freedom to choose the other $M$ eigenvectors of $\mathcal{T}$, and so there could be non-classical dynamics.
\item {\em Fully independent theories} have $N+M=d$ eigenvectors with eigenvalue $+1$. This completely fixes $\mathcal{T}=\id$, and so there can not be any non-classical dynamics.
\end{enumerate}
\end{proof}

Note that from this derivation, we can also make a broader statement for theories that are only partially conditionally restricted: {\em each independent conditional restriction of state-space we place grants us more freedom in our choice of transformation}. This follows naturally from the fact that placing a restriction on the state-space prevents us from selecting the full set of $N+M-1$ independent eigenvectors from the branch locality restriction.


\begin{thebibliography}{10}

\bibitem{Heisenberg30}
W.~K. Heisenberg,
\newblock {\em The physical principles of the quantum theory} (Dover, New York,
  NY, 1930).

\bibitem{Bohr58}
N.~Bohr,
\newblock American Journal of Physics {\bf 26} (1958).

\bibitem{PopescuR94}
S.~Popescu and D.~Rohrlich,
\newblock Found. Phys. {\bf 24}, 379 (1994).

\bibitem{Barrett07}
J.~Barrett,
\newblock \pra {\bf 75}, 032304 (2007).

\bibitem{VerSteegW08}
G.~verSteeg and S.~Wehner,
\newblock QIC {\bf 9}, 0801 (2009).

\bibitem{OppenheimW10}
J.~Oppenheim and S.~Wehner,
\newblock Science 19 {\bf 330}, 1072 (2010), arXiv:1004.2507.

\bibitem{HanggiW13}
E.~H\"{a}nggi and S.~Wehner,
\newblock Nature communications {\bf 4}, 1670 (2013).

\bibitem{Hardy01}
L.~{Hardy},
\newblock (2001), arXiv:quant-ph/0101012.

\bibitem{Mana03}
P.~{Mana},
\newblock arXiv:quant-ph/0305117v3  (2003).

\bibitem{BarnumBLW06}
H.~Barnum, J.~Barrett, M.~Leifer, and A.~Wilce,
\newblock \prl {\bf 99}, 240501 (2007).

\bibitem{DakicB11}
B.~{Dakic} and C.~{Brukner},
\newblock Deep Beauty(...) Ed. H. Halvorson, CUP , 365 (2011).

\bibitem{MasanesM11}
L.~{Masanes} and M.~P. {M\"uller},
\newblock NJP {\bf 13}, 063001 (2011).

\bibitem{MullerDV11}
M.~P. M\"uller, O.~C.~O. Dahlsten, and V.~Vedral,
\newblock Communications in Mathematical Physics {\bf 316}, 441 (2012).

\bibitem{GarnerDNMV13}
A.~J.~P. Garner, O.~C.~O. Dahlsten, Y.~Nakata, M.~Murao, and V.~Vedral,
\newblock To appear in NJP  (2013), arXiv:quant-ph/1304.5977.

\bibitem{Sorkin94}
R.~Sorkin,
\newblock Mod. Phys. Lett. A {\bf 9}, 3119 (1994).

\bibitem{BarnettDR08}
M.~Barnett, F.~Dowker, and D.~Rideout,
\newblock J. Phys. A: Math. Theor. {\bf 40}, 7255 (2007).

\bibitem{UdudecBE11}
C.~Ududec, H.~Barnum, and J.~Emerson,
\newblock Found Phys {\bf 41}, 15 (2011),
\newblock {a}rXiv:1003.5005.

\bibitem{ShortB10}
A.~J. Short and J.~Barrett,
\newblock New Journal of Physics {\bf 12}, 033034 (2010).

\bibitem{GrossMCD10}
D.~Gross, M.~M\"uller, R.~Colbeck, and O.~C.~O. Dahlsten,
\newblock \prl {\bf 104}, 080402 (2010).

\bibitem{DeutschJ92}
D.~{Deutsch} and R.~{Jozsa},
\newblock Proc. R. Soc. Lond. A. {\bf 439}, 553 (1992).

\bibitem{CleveEMM98}
R.~{Cleve}, A.~{Ekert}, C.~{Macchiavello}, and M.~{Mosca},
\newblock Proc. R. Soc. Lond. A. {\bf 454}, 339 (1998).

\bibitem{Zeilinger99}
A.~{Zeilinger},
\newblock Found.Phys. {\bf 29} (1999).

\bibitem{PaterekDB10}
T.~Paterek, B.~Dakic, and C.~Brukner,
\newblock NJP {\bf 12}, 053037 (2010).

\bibitem{BruknerZ09}
C.~Brukner and A.~Zeilinger,
\newblock Found. Phys. {\bf 39}, 677 (2009).

\end{thebibliography}
\end{document}